\documentclass[12pt]{iopart}
\usepackage{graphicx}	
\usepackage{bm}

\newcommand{\eqref}[1]{(\ref{#1})}
\renewcommand{\vec}[1]{\bm #1}

\begin{document}
\title{
Optimized effective potential method
and application to static RPA correlation
}
\author{Taro Fukazawa}
\address{
   Graduate School of Engineering
   Science, Osaka University
}
\author{Hisazumi Akai}
\address{
Institute for Solid State Physics, University of Tokyo
}
\begin{abstract}
The optimized effective potential (OEP) method is 
a promising technique for calculating the ground state properties
of a system within the density functional theory. 
However,
it is not widely used
as its computational cost is rather high and, also, some
ambiguity remains in the theoretical framework.
In order to overcome these problems,
we first introduced a method that accelerates the OEP scheme in a static
RPA-level correlation functional. 
Second, the
Krieger-Li-Iafrate (KLI) approximation is exploited
to solve the OEP equation.
Although seemingly too crude, this approximation did not
reduce the accuracy of 
the description of the magnetic transition metals (Fe, Co, and Ni) examined here,
the magnetic properties of which are rather sensitive to
correlation effects.
Finally, we reformulated the OEP method to render it applicable
to the direct RPA correlation functional
and other, more precise, functionals.
Emphasis is placed on the following three points of the discussion:
i) Level-crossing at the Fermi surface is taken into account;
ii) eigenvalue variations in a Kohn-Sham functional
are correctly treated; and
iii) the resultant OEP equation is different from those reported
to date.

\end{abstract}

\submitto{\JPCM}
\maketitle

\section{Introduction}
In the Kohn-Sham scheme of density functional
theory, 
the exchange-correlation energy functional
plays a key role in determining
the accuracy of a calculation
because
the majority of the many-body effects are incorporated in
the functional.
The local density
approximation (LDA) has been the most widely-used approximation
of this functional,
as a result of its simplicity, small computational cost,
and its unexpectedly accurate description of matter.
However, there is a growing need
for methods with both higher accuracy and
tolerable computational cost for
practical use. 
The optimized effective potential (OEP)
method\cite{SH:PhysRev.90.317,PhysRevA.14.36}
is a promising approach that 
offers
a means of including more accurate exchange-correlation functionals in such calculations.
Its theoretical flexibility comes from the fact that 
the OEP is a general method used to minimize the energy functionals that
are explicitly expressed in terms of the Kohn-Sham orbitals.
Of the possible functionals,
a formalism that takes account of the random-phase-approximation (RPA)-level
correlation is
considered to be 
an important step
as regards reasonable descriptions of solid systems.
Kotani\cite{kotani98} included such RPA-level correlation
with a static approximation and obtained reasonable results
for realistic systems.
These results were sufficiently encouraging,
however,
this type of self-consistent calculation has not become 
widely-used.
This can be attributed to its highly
time-consuming procedures, which are used during every
iteration step in order to obtain a self-consistent result.

Two of the procedures are rate-determining.
One, which is common to all the OEP calculations, is
a procedure that solves the OEP equation.
As regards this stage,
several authors have already proposed
techniques\cite{PhysRevLett.90.043004,PhysRevB.68.035103,
kosaka2006simple,PhysRevB.85.235126}
and approximations\cite{Krieger1990256,KLI:PhysRevA.45.101, 
staroverov2006optimized,staroverov2006effective,
izmaylov2007effective,izmaylov2007self,PhysRevA.64.042506,
della2001efficient}
that
accelerate the process.
The other rate-determining procedure is the 
calculation of the variation of the static RPA (sRPA) correlation energy,
which has attracted only minor attention to date.
However, this is an unavoidable issue if one considers
practical use of the sRPA functional.
For the former procedure, 
an approximation
of the OEP equation
proposed by Krieger, Li, and
Iafrate\cite{Krieger1990256,KLI:PhysRevA.45.101}
(KLI) will be used in the present paper.
It appears to be a crude approximation, and 
it was also shown by Krieger et al. that
this formalism can be regarded as only a lowest level 
approximation in a 
systematic improvement scheme\cite{krieger1992systematic}.
However, it has been reported that the KLI approximation does not 
contradict the numerical predictions for some isolated
systems\cite{Krieger1990256,KLI:PhysRevA.45.101,grabo1995density}.
We also reported\cite{fukazawa2010new}
that it does not negatively affect descriptions of some extended
systems within the exact-exchange (EXX) functional level,
provided a slight modification to the KLI scheme is employed.
Then, a natural question is 
whether the modified KLI approximation
is still feasible within the OEP+sRPA scheme.
In order to answer this query, errors from additional procedures used
to calculate the correlation effects must be suppressed.
For example, the
summation of unoccupied orbitals in the procedure for calculating
the variation of the sRPA functional seems a considerable
obstacle to retaining the necessary accuracy;
this also renders the procedure time-consuming.
In order to achieve both acceleration and precision,
we have devised a method that
avoids such a summation
without loss of accuracy.
We will show that, with this method, the modified KLI approximation does
not reduce the accuracy of
the descriptions of the magnetic transition metals studied here, namely, Fe, Ni, and Co.

Reformulation of the OEP method is also discussed in the present paper.
In our opinion, the standard derivations of 
the OEP equation are incomplete.
This is reflected in an unphysical degree of freedom entering
the equation obtained by
Sharp and Horton\cite{SH:PhysRev.90.317}
and by Talman and Shadwick\cite{PhysRevA.14.36} (SHTS).
In our discussion, we will incorporate the dependency of
the Kohn-Sham energy-functional on
eigenvalues and also consider level-crossing at the Fermi surface.
In addition, we will discuss how these factors have been neglected in the standard
derivations.
Concerning the level-crossing effects, 
we have already presented some of this discussion 
in our previous paper\cite{fukazawa2010new}, with a view to providing a complementary
equation to the SHTS equation.
A result in the present paper justifies the use of this complementary
equation.
As for treatment of the eigenvalues, this has been ignored for a long
time.
However, variations with respect to the eigenvalues cannot be ignored in these calculations, 
because they are linked to a variation of density via an effective potential
in the Kohn-Sham equation, with
\begin{equation}
 \left[ -\nabla^2 + V_{{\rm eff},\sigma}(\vec r) \right]
 \psi_{i\sigma}(\vec r)
 =
 \epsilon_{i\sigma}  \psi_{i\sigma}(\vec r),
 \label{KSeq1}
\end{equation}
where the ground-state density determines the corresponding effective
potential, $V_{\rm eff}$. The Kohn-Sham orbital, $\psi_{i\sigma}$,
and its eigenvalue, $\epsilon_{i\sigma}$,
are determined by $V_{\rm eff}$.
With consideration of the eigenvalue-dependency,
we will derive a correction term to the
SHTS equation in the following section.
It is also shown that this correction term is fortunately canceled
in the static RPA functional,
but this cancellation does not occur in general.

This paper will show the derivation of OEP-EXX+sRPA technique,
reformulate the OEP method, and provide sample calculations and results.
The structure of the paper is as follows:
Section \ref{SS.Theory} is devoted to our
formalism.
In this section, for convenience, reformulation of the OEP precedes specific topics
concerning the sRPA
correlation.
Details and results of the calculations obtained using our OEP-EXX+sRPA technique
are shown and discussed
in section \ref{SS.Results}.
In the last section, section \ref{SS.Conclusion},
we present a summary and 
conclusions.

\section{Theory}
\label{SS.Theory}

\subsection{Reformulation of optimized effective potentials}
In the Kohn-Sham scheme, the ground-state spin-density,
$n_{\sigma}^{0}$,
is
determined through
minimization of the Kohn-Sham energy functional, $E_{\rm KS}$, such that
\begin{equation}
 \left.
    \frac
    {
       \delta E_{\rm KS}
       [
          n_{\uparrow},
          n_{\downarrow}
       ]
    }
    {\delta n_{\sigma}({\vec r})}
 \right|_{n_{\sigma}=n_{\sigma}^0}
 = 0,
\label{HKvar}
\end{equation}
where the number of electrons 
and the spin-density,  $n_{-\sigma}$, of the opposite spin direction is
fixed.
In the standard derivations of the OEP equation, the variation of the
density, $n_{\sigma}$, is decomposed into the variations of the effective
potential,
$V_{{\rm eff},\sigma}$, the Kohn-Sham orbitals, $\{\psi_{i\sigma}\}_i$,
and their conjugates, with
\begin{equation}
 \frac{\delta}{\delta n_{\sigma}(\vec r)}
 =
 \sum_{i}
 \int d{\vec r}'
 \,   d{\vec r}''
 \left(
 \frac{\delta V_{{\rm eff},\sigma}({\vec r}')}{\delta n_{\sigma}({\vec r})}
 \frac{\delta \psi_{i\sigma}({\vec r}'')}
      {\delta V_{{\rm eff},\sigma}({\vec r}')}
 \frac{\delta}{\delta \psi_{i\sigma}({\vec r}'')}
 + 
 \rm{c.c.}
 \right).
 \label{FalseChain}
\end{equation}
However, it is often forgotten that
the variation of the effective potential in
equation \eqref{FalseChain}
must be restricted in an unusual way in order to keep
$n_{-\sigma}$ fixed.
Specifically, variation of the potential ($\delta V_{{\rm eff},\sigma}$) that 
perturbs an
eigenvalue, $\epsilon_{i\sigma}$, away from the Fermi sea
must be excluded if an unoccupied opposite-spin state becomes  
occupied. This is in order to conserve 
the number of electrons in the perturbed system.
Otherwise, a change in $n_{-\sigma}$ is permitted.
Neglecting this restriction on $\delta V_{{\rm eff},\sigma}$
leads to
an incorrect value for
$
\frac{\delta V_{{\rm eff},\sigma}}
     {\delta n_{\sigma}}
$
in equation \eqref{FalseChain},
and an extra degree of freedom
is introduced to the resultant potential.
This explains how the well-known unphysical insensitivity
to the transform
\begin{equation}
 V_{{\rm eff},\sigma} \rightarrow V_{{\rm eff},\sigma}+C_{\sigma}
\label{unPhysFreedom}
\end{equation}
is introduced in the SHTS equation,
where $C_{\sigma}$ is an arbitrary real number.

There appears to be no easy means of removing such a forbidden
$\delta V_{{\rm eff},\sigma}$
in
$
\frac{\delta V_{{\rm eff},\sigma}}
     {\delta n_{\sigma}}.
$
However, 
this difficulty
can be avoided by taking $V_{{\rm eff},\uparrow}$ and 
$V_{{\rm eff},\downarrow}$  as  variables independent from each other
in the minimization of $E_{\rm KS}$, and by allowing
$V_{{\rm eff},\sigma}$ to change $n_{-\sigma}$.
However, in this alternative strategy, the implicit expectation that
$E_{\rm KS}$ is a functional of Kohn-Sham orbitals only, as described
in equation \eqref{FalseChain}, is no longer acceptable,
because $V_{{\rm eff},\sigma}$ can change $n_{-\sigma}$ 
by only perturbing
eigenvalues of spin-direction $\sigma$.
Let us take the LDA for example.
The LDA can be treated in the standard OEP derivation with the relation
\begin{equation}
 n_{\sigma}(\vec r)[\{\psi^*_{i\sigma},\psi_{i\sigma}\}_i]
 =
 \sum_{i=1}^{N_{\sigma}}
 \psi^{*}_{i\sigma}(\vec r)
 \psi_{i\sigma}(\vec r),
\label{traditionaldensity}
\end{equation}
where $N_{\sigma}$ denotes the number of occupied Kohn-Sham orbitals
for spin $\sigma$.
Here, the Kohn-Sham functional
should be expressed
explicitly in terms of orbitals
as
\begin{equation}
E^{\rm LDA}_{\rm KS}
[\{\psi^*_{i\sigma},\psi_{i\sigma}\}]
=
J_{\rm KS}[\{\psi^*_{i\sigma},\psi_{i\sigma}\}]
+
E^{\rm LDA}_{\rm xc}
[\{n_{\sigma}[\{\psi^*_{i\sigma},\psi_{i\sigma}\}_i]\}],
\label{traditionalLDA}
\end{equation}
with
\begin{equation}
J_{\rm KS}
=\sum_{\sigma}\sum_{i=1}^{N_{\sigma}}
\int d{\vec r}\,
\psi^*_{i\sigma}(\vec r)
[-\nabla^2+V_{\rm ext}(\vec r)]
\psi_{i\sigma}(\vec r)
+ E_{\rm Hartree}[\{\psi^*_{i\sigma},\psi_{i\sigma}\}],
\end{equation}
and 
\begin{equation}
E_{\rm Hartree}
=
\sum_{\sigma,\sigma'}
\sum_{i=1}^{N_{\sigma}}
\sum_{j=1}^{N_{\sigma'}}
\int d{\vec r}\, d{\vec r}' \,
\frac{
\psi^*_{i\sigma}(\vec r)
\psi_{i\sigma}(\vec r)
\psi^*_{j\sigma'}({\vec r}')
\psi_{j\sigma'}({\vec r}')
}
{\left|
{\vec r}-{\vec r}'
\right|}.
\end{equation}
In a standard treatment, $N_{\sigma}$ must be fixed during the variation
because $N=N_{\uparrow}+N_{\downarrow}$ must be conserved and 
$n_{-\sigma}$ must be constant.
In contrast, we must treat the implicit
eigenvalue-dependence of $N_{\sigma}$ in our alternative
strategy
because we are allowing 
$V_{{\rm eff},\sigma}$
to change $N_{-\sigma}$ through eigenvalues at the Fermi level.

However, one should also note that
a set of eigenvalues, $\{\epsilon_{i\sigma}\}$, is not
a functional of
the spin-density only
because there exists a freedom in the choice of reference energy.
On the other hand, the exact $E_{\rm KS}$ is a functional of
the spin-density only.
Therefore, it is natural to introduce 
another variable that can refer to the origin of the energy,
so that it can cancel unwanted dependences on the reference energy.
Let $\mu$ denote such a variable,
one of the possible definitions of which is as follows.
For a given $\{n_{\sigma}\}$, the independent variable $\mu$ 
determines $\{\epsilon_{i\sigma}\}$
so that
$\mu$ becomes a minimum real number 
that causes the relation
\begin{equation}
 N
  =
  \lim_{\delta\mu\rightarrow 0^+}
  \sum_{i,\sigma}
  \theta(\mu+\delta\mu-\epsilon_{i\sigma})\ ,
\label{number-electron}
\end{equation}
to hold, where $\theta$ is the step function
and the summation is taken over 
all the orbitals regardless of their occupancy.
In order for $\{\epsilon_{i\sigma}\}$ to be determined
uniquely, even for cases involving semiconductors or insulators,
the ``minimum real'' feature is required.
In most cases, one can neglect the existence of $\delta \mu$ by 
simply substituting zero in its place and defining $\theta(0)=1$.
We use this condition for simplicity hereafter.
Note that
the dependency of each $\epsilon_{i\sigma}$ on the energy-reference is
canceled 
by subtraction of $\mu$ in equation \eqref{number-electron}.
Let us return to the LDA example 
with the definition of $\mu$ given above.
The spin-density, $n_{\sigma}$, can be expressed naturally using $\mu$ as
\begin{equation}
 n_{\sigma}({\vec r})
  =
  \sum_{i}
  \theta(\mu-\epsilon_{i\sigma})\,
  \psi^*_{i\sigma}({\vec r})
  \psi_{i\sigma}({\vec r}),
\label{spin-density0}
\end{equation}
and 
is now re-expressed as a functional of
$\{\psi^*_{i\sigma},\psi_{i\sigma}\}$,
$\{\epsilon_{i\sigma}\}$, and $\mu$.

Equation \eqref{traditionalLDA} in its entirety can also be given 
naturally as
\begin{eqnarray}
 E^{\rm LDA}_{\rm KS}
 \left[
   \{\psi^*_{i\sigma},\psi_{i\sigma}\},
   \{\epsilon_{i\sigma}\}
 ,\mu
 \right] \nonumber \\
 =
 J_{\rm KS}
 \left[
   \{\psi^*_{i\sigma},\psi_{i\sigma}\},
   \{\epsilon_{i\sigma}\}
 ,\mu
 \right]
 +
 E^{\rm LDA}_{\rm xc}
 \left[
 \left\{
   n_{\sigma}
   \left[
     \{\psi^*_{i\sigma},\psi_{i\sigma}\}_i,
     \{\epsilon_{i\sigma}\}_i
   ,\mu
   \right]
 \right\}_{\sigma}
 \right],
\label{LDAnew}
\end{eqnarray}
with 
\begin{equation}
 J_{{\rm KS}}
 =
 \sum_{i,\sigma}
 \theta(\mu-\epsilon_{i\sigma})\,
 \psi^*_{i\sigma}({\vec r})
 \left[
   -\nabla^2
   +
   V_{\rm ext}({\vec r})
 \right]
 \psi_{i\sigma}({\vec r})
 +
 E_{\rm Hartree}
 \left[
   \{\psi^*_{i\sigma},\psi_{i\sigma}\},
   \{\epsilon_{i\sigma}\}
 ,\mu
 \right],
\label{JKS1}
\end{equation}
and 
\begin{equation}
 E_{\rm Hartree}
 =
 \sum_{i,\sigma}
 \theta(\mu-\epsilon_{i\sigma})
 \sum_{j,\sigma'}
 \theta(\mu-\epsilon_{j\sigma'})
 \int d{\vec r}\, d{\vec r}' \,
 \frac{
 \psi^*_{i\sigma}({\vec r})
 \psi_{i\sigma}({\vec r})
 \psi^*_{j\sigma'}({\vec r}')
 \psi_{j\sigma'}({\vec r}')
 }
 {\left|{\vec r}-{\vec r}'\right|},
\end{equation}
where $V_{\rm ext}$ is a fixed external potential.

As shown in the above example, we must manage a functional that
depends at
least on
$\{\psi^*_{i\sigma},\psi_{i\sigma}\}$,
$\{\epsilon_{i\sigma}\}$, and $\mu$ in general.
Therefore, 
in the following, we consider a generalized Kohn-Sham functional 
in the form,
$
 E_{\rm KS}
 \left[
   \{\psi^*_{i\sigma},\psi_{i\sigma}\},
   \{\epsilon_{i\sigma}\}
 ,\mu
 \right].
$
Equations for
$V_{{\rm eff},\uparrow}$, $V_{{\rm eff},\downarrow}$,
and $\mu$
are obtained by minimizing $E_{\rm KS}$ with the number of electrons fixed
and the ``minimum real'' requirement for $\mu$.
Each of the variables,
$V_{{\rm eff},\uparrow}$, $V_{{\rm eff},\downarrow}$,
and $\mu$,
can be treated as being independent 
from each other through use of the Lagrange multiplier method.
Note that the generalized $E_{\rm KS}$ includes functionals
that
depend on
the choice of reference energy as well as 
those that do not.
The properties of the Kohn-Sham functionals that are invariant
under 
alteration of the reference energy will be addressed later.
%

We now discuss
the minimization of $E_{\rm KS}$, again
adopting
the definition of $\mu$ given in \eqref{number-electron}.
The ``minimum real'' requirement for $\mu$ in
equation \eqref{number-electron}
is expressed in an analytic
form, such that
\begin{equation}
P=
\lim_{\delta\mu \rightarrow 0^+}
\prod_{i\sigma} \theta(\mu+\delta\mu-\epsilon_{i\sigma})
\times (\mu+\delta\mu-\epsilon_{i\sigma})
=0, \\
\end{equation}
which is sufficient because
a constraint on
equation \eqref{number-electron} is also required for electron-number conservation.
The Kohn-Sham functional can be 
written as 
\begin{equation}
 E_{\rm KS}
 \left[
   \{\psi^*_{i\sigma},\psi_{i\sigma}\},
   \{\epsilon_{i\sigma}\}
 ,\mu
 \right]
 =
 J_{\rm KS}
 \left[
   \{\psi^*_{i\sigma},\psi_{i\sigma}\},
   \{\epsilon_{i\sigma}\}
 ,\mu
 \right]
 +
 E_{\rm xc}
 \left[
     \{\psi^*_{i\sigma},\psi_{i\sigma}\},
     \{\epsilon_{i\sigma}\},
     \mu
 \right],
\label{generalKS}
\end{equation}
with $J_{\rm KS}$ as defined in equation \eqref{JKS1}.
The chain rule for the variation of $V_{{\rm eff},\sigma}({\vec r})$ is 
\begin{equation}
 \frac
 {\delta}
 {\delta V_{{\rm eff},\sigma}({\vec r})}
 =
 \sum_i
 \left[
    \int
    \left\{
    \frac
       {\delta \psi_{i\sigma}({\vec r}')}
       {\delta V_{{\rm eff},\sigma}({\vec r})}
    \frac
       {\delta}
       {\delta \psi_{i\sigma}({\vec r}')}
    +
    \frac
       {\delta \psi^*_{i\sigma}({\vec r}')}
       {\delta V_{{\rm eff},\sigma}({\vec r})}
    \frac
       {\delta}
       {\delta \psi^*_{i\sigma}({\vec r}')}
    \right\}
    d{\vec r}'
    +
    \frac
       {\delta \epsilon_{i\sigma}}
       {\delta V_{{\rm eff},\sigma}({\vec r})}
    \frac
       {\partial}
       {\partial \epsilon_{i\sigma}}
 \right].
\end{equation}
Hence, one obtains from the variation with respect to $V_{\rm eff}$
\begin{eqnarray}
 \sum_{i}
  \int
  d{\vec r}'
  \left\{
  \left[
     \theta(\mu-\epsilon_{i\sigma})
     V_{{\rm xc},\sigma}({\vec r}')
     -
     v_{{\rm xc},i\sigma}({\vec r}')
  \right]
  \mathbf{G}_{i\sigma}({\vec r}',{\vec r})
  \psi_{i\sigma}^*({\vec r}')
  \psi_{i\sigma}({\vec r})
  +
  {\rm c.c.}
  \right\} \nonumber \\
  -
 \sum_{i}
 \left[
  \delta (\mu-\epsilon_{i\sigma})
   \int
   \left\{
  \xi+\mu-V_{{\rm xc},\sigma}({\vec r}')
   \right\}
  \psi_{i\sigma}^*({\vec r}')
  \psi_{i\sigma}({\vec r}')
  d{\vec r}'
  -
    \frac
       {\partial {E}_{\rm xc}}
       {\partial \epsilon_{i\sigma}}
  - \lambda
    \frac
       {\partial {P}}
       {\partial \epsilon_{i\sigma}}
  \right]
  \psi_{i\sigma}^*({\vec r})
  \psi_{i\sigma}({\vec r})
  \nonumber \\
  =
  0,
  \label{NewMainEq}
\end{eqnarray}
where 
\begin{eqnarray}
 \mathbf{G}_{i\sigma}({\vec r}',{\vec r})
  =
  \sum_{\{j|\epsilon_{i\sigma}\neq\epsilon_{j\sigma}\}}
  \frac
  {
     \psi_{j\sigma}^*({\vec r})
     \psi_{j\sigma}({\vec r}')
  }
  {\epsilon_{i\sigma}-\epsilon_{j\sigma}}, \\
 v_{{\rm xc},i\sigma}
  =
  \frac{1}{\psi_{i\sigma}^*({\vec r})}
  \frac{\delta {E}_{\rm xc}}{\delta \psi_{i\sigma}({\vec r})}.
\end{eqnarray}
Here, $\xi$ and $\lambda$ denote the Lagrange multipliers for the constraint
on the number of electrons and for $P=0$, respectively,
and $V_{{\rm xc},\sigma}$ is 
the exchange-correlation potential (defined as 
$V_{{\rm xc},\sigma}=V_{{\rm eff},\sigma}-V_{\rm ext}-V_{\rm Hartree}$).
It should be noted that
equation \eqref{NewMainEq} appears to be in a form that is not invariant
under unitary mixing among orbitals
belonging to a degenerate level.
This strange feature comes from
the application of the elementary perturbation theory
in the derivation, specifically, from
$
\frac{\delta \epsilon_{i\sigma}}
     {\delta V_{{\rm eff},\sigma}(\vec r)}
$.
In the treatment of the perturbation,
degenerate orbitals must be arranged
by a unitary transform
so that the perturbed orbitals also become
eigenstates of the perturbed system.
Although the appropriate choice of unitary transform depends on
the perturbation in question, 
recall that $\epsilon_{i\sigma}$ transforms as a 
diagonal element of a tensor, $\epsilon_{ij,\sigma}$, under a unitary
transform when $\sigma$ is fixed, and that 
we have chosen a convenient transform so that no off-diagonal element
of $\epsilon_{ij,\sigma}$ is found in 
 $J_{\rm KS}$, $E_{\rm xc}$, and $P$. Thus, 
equation \eqref{NewMainEq} can be rewritten as 
\begin{eqnarray}
 \sum_{i}
  \int
  d{\vec r}'
  \left\{
  \left[
     \theta(\mu-\epsilon_{i\sigma})
     V_{{\rm xc},\sigma}({\vec r}')
     -
     v_{{\rm xc},i\sigma}({\vec r}')
  \right]
  \mathbf{G}_{i\sigma}({\vec r}',{\vec r})
  \psi_{i\sigma}^*({\vec r}')
  \psi_{i\sigma}({\vec r})
  +
  {\rm c.c.}
  \right\} \nonumber \\
  +
 \sum_{ij}
 \left[
    \frac
       {\partial {J}_{\rm KS}}
       {\partial \epsilon_{ij,\sigma}}
    +
    \frac
       {\partial {E}_{\rm xc}}
       {\partial \epsilon_{ij,\sigma}}
  + \lambda
    \frac
       {\partial {P}}
       {\partial \epsilon_{ij,\sigma}}
  \right]
  \psi_{j\sigma}^*({\vec r})
  \psi_{i\sigma}({\vec r})
  =
  0,
\label{halfUnitary}
\end{eqnarray}
which is invariant under unitary transforms.
This proves the validity of equation \eqref{halfUnitary} under any unitary
transform,
and it follows that equation \eqref{NewMainEq} also holds for any
unitary transform that diagonalizes $\epsilon_{ij,\sigma}$.
Therefore, the necessity of selecting an appropriate
unitary transform according to the perturbation theory
in the derivation of \eqref{NewMainEq} is no longer pertinent.

Equation \eqref{NewMainEq} can be decomposed into two equations
and two identities can be obtained,
$A=0$ and $B=0$,
from $\delta(0)\times A+B=0$ and vice versa.
In order to achieve this, we separate
$
 \frac
  {\partial {E}_{\rm xc}}
  {\partial \epsilon_{i\sigma}}
$
into two parts, one denoted by
${D}_{{\rm xc},{i\sigma}}$,
which includes all the terms proportional to
$\delta(\mu-\epsilon_{i\sigma})$,
with 
${R}_{{\rm xc},{i\sigma}}$
representing the remaining terms. Hence,
\begin{equation}
 \frac
  {\partial {E}_{\rm xc}}
  {\partial \epsilon_{i\sigma}}
 =
 {D}_{{\rm xc},i\sigma}
 +
 {R}_{{\rm xc},i\sigma}.
\label{DecomEe}
\end{equation}
We also assume that $E_{\rm xc}$ does not have a singularity in itself,
so there is no
$\delta$-function in the form of
$\delta(\mu-\epsilon_{j\sigma'})$ 
in ${R}_{{\rm xc},i\sigma}$.
Then, the following equations are obtained by substituting
equation \eqref{DecomEe} into equation \eqref{NewMainEq}:
\begin{eqnarray}
 \sum_{i}
  \int
  d{\vec r}'
  \left\{
  \left[
     \theta(\mu-\epsilon_{i\sigma})
     V_{{\rm xc},\sigma}({\vec r}')
     -
     v_{{\rm xc},i\sigma}({\vec r}')
  \right]
  \mathbf{G}_{i\sigma}({\vec r}',{\vec r})
  \psi_{i\sigma}^*({\vec r}')
  \psi_{i\sigma}({\vec r})
  +
  {\rm c.c.}
  \right\} \nonumber \\
 \quad
  +
  \sum_{i}
  \left(
  {R}_{{\rm xc},i\sigma}
  +
  \lambda
  \frac{\partial P}{\partial \epsilon_{i\sigma}}
  \right)
  \psi_{i\sigma}^*({\vec r})
  \psi_{i\sigma}({\vec r})
  =
  0,
  \label{ModSHTS}
\end{eqnarray}
and 
\begin{equation}
 \sum_i
 \left[
  \delta (\mu-\epsilon_{i\sigma})
   \int
   \left\{
  \xi+\mu-V_{{\rm xc},\sigma}({\vec r}')
   \right\}
  \psi_{i\sigma}^*({\vec r}')
  \psi_{i\sigma}({\vec r}')
  d{\vec r}'
  -
  {D}_{{\rm xc},i\sigma}
  \right]
  \psi_{i\sigma}^*({\vec r})
  \psi_{i\sigma}({\vec r})
  =
  0,
  \label{OnWayFA}
\end{equation}
where terms proportional to the $\delta$-function in derivatives of $P$ vanish
because $P=0$ holds.

At first glance, equation \eqref{OnWayFA} may seem to imply that the expression in the square
bracket should be
zero for each $i$.
This is not true, however, as will be shown below.
Consider a $\mathcal{N} \times \mathcal{N}$ matrix whose $i,j$ element
is given by
$\psi^{*}_{i\sigma}({\vec r})\psi_{j\sigma}({\vec r})$, where the labels
are taken so that 
all the orbitals with $\sigma$ spin at the highest occupied level are
given by
$\psi_{1\sigma}({\vec r}), \cdots, \psi_{\mathcal{N}\sigma}({\vec r})$.
This matrix, if diagonalized, 
has only one non-zero diagonal element.
Therefore, the diagonal element
$\psi^{*}_{i\sigma}\psi_{i\sigma}$ of this matrix, appearing next to the
bracket in equation \eqref{OnWayFA}, contains no more information than a
single scalar.
This is the reason why one cannot derive relations greater than 1 from
equation \eqref{OnWayFA}.
It is also notable that all the diagonal elements 
can always be made identical using a unitary transform that maps
$(\psi_{1\sigma}({\vec r}), \cdots, \psi_{\mathcal{N}\sigma}({\vec r}))$ 
to a vector proportional to
$(1, \cdots, 1)$. With this choice of the bases, equation
\eqref{OnWayFA} becomes
\begin{equation}
  \sum_{i}
  \left[
  \delta (\mu-\epsilon_{i\sigma})
   \int
   \left\{
  \xi+\mu-V_{{\rm xc},\sigma}({\vec r}')
   \right\}
  \psi_{i\sigma}^*({\vec r}')
  \psi_{i\sigma}({\vec r}')
  d{\vec r}'
  -
  {D}_{{\rm xc},i\sigma}
  \right]
  =
  0.
\label{OnWayFA3}
\end{equation}
This equation is again in a form invariant under any unitary mixing
among the highest occupied orbitals, which proves that
equation \eqref{OnWayFA3} holds regardless of the choice of bases,
provided equation
\eqref{OnWayFA} holds. Conversely, equation \eqref{OnWayFA} 
can be derived from \eqref{OnWayFA3} in a similar way.
Therefore, \eqref{OnWayFA} and \eqref{OnWayFA3} are equivalent to 
each other. To summarize, we have obtained the following equivalence
between equations
\begin{equation}
 \eqref{NewMainEq} \Longleftrightarrow \eqref{ModSHTS},\eqref{OnWayFA3}.
 \label{equiv0}
\end{equation}

As for the variation of $E_{\rm KS}$
with respect to
$\mu$, this
yields
\begin{equation}
\sum_{i,\sigma}
 \left[
  \delta (\mu-\epsilon_{i\sigma})
   \int
   \left\{
  \xi+\mu-V_{{\rm xc},\sigma}({\vec r}')
   \right\}
  \psi_{i\sigma}^*({\vec r}')
  \psi_{i\sigma}({\vec r}')
  d{\vec r}'
  \right] 
  +
    \frac
       {\partial {E}_{\rm xc}}
       {\partial \mu}
  +
  \lambda
  \frac{\partial P}{\partial \mu}
 =
 0.
\label{FAequation}
\end{equation}
Noting that integration of \eqref{ModSHTS} with respect to
$\vec r$ gives
\begin{equation}
 \sum_{i\sigma}
 R_{{\rm xc},i\sigma}
 =
 -
 \sum_{i\sigma}
 \lambda
 \frac{\partial P}
      {\partial \epsilon_{i\sigma}}
 =
 \lambda
 \frac{\partial P}
      {\partial \mu},
 \label{Rsum}
\end{equation}
it is straightforward to show the following equivalency
\begin{equation}
 \eqref{NewMainEq}, \eqref{FAequation}
 \Longleftrightarrow
 \eqref{NewMainEq}, \eqref{convcond}
 \label{equiv2} 
\end{equation}
where equation \eqref{convcond} is 
\begin{equation}
  \frac
    {\partial {E}_{\rm KS}}
    {\partial \mu}
 +
 \sum_{i\sigma}
 \frac
    {\partial {E}_{\rm KS}}
    {\partial \epsilon_{i\sigma}}
 =
 0.
\label{convcond}
\end{equation}
This equation has an important  meaning as regards
the dependency of $E_{\rm KS}$ on the reference energy.
As mentioned above,
we introduced the variable $\mu$ so that it could be possible to cancel the 
artificial dependency.
However, we did not exclude the possibility that 
$E_{\rm KS}$ depends on the energy-scale reference
in the above discussion, such that
\begin{equation}
 E_{\rm KS}
  [n_{\uparrow},n_{\downarrow},\mu]
 =
 {E}_{\rm KS}
  [
    \{
       \{\psi^*_{i\sigma},\psi_{i\sigma}\}_i[n_{\sigma}]
    \}_{\sigma},
    \{\epsilon_{i\sigma}\}[n_{\uparrow},n_{\downarrow},\mu],
    \mu
  ].
\label{HKconnection_check}
\end{equation}
When $E_{\rm KS}$ does not depend on $\mu$, the total derivative of the
right-hand side with respect to $\mu$ is always zero, which can 
be expressed in exactly the same manner as equation \eqref{convcond}.
However, the above general discussion requires
equation \eqref{convcond} only
at extrema of the functional.

To distinguish the special case from the general discussion, 
we introduce $\dot{E}_{\rm KS}$ to denote a Kohn-Sham functional that
always satisfies \eqref{convcond}.
The equivalency indicated in \eqref{equiv2} ensures that 
the partial derivative of
$\dot{E}_{\rm KS}[
       \{\psi^*_{i\sigma},\psi_{i\sigma}\},
       \{\epsilon_{i\sigma}\},
    \mu
]$
with respect to $\mu$ does not require consideration.
%
The reference-energy-free
functional can also be expressed as
\begin{equation}
 \dot{E}_{\rm KS}
  [n_{\uparrow},n_{\downarrow}]
 =
 \dot{E}_{\rm KS}
  [
    \{
       \{\psi^*_{i\sigma},\psi_{i\sigma}\}_i[n_{\sigma}]
    \}_{\sigma},
    \{\varepsilon_{i\sigma}\}[n_{\uparrow},n_{\downarrow}]
  ],
\end{equation}
where the variables $\{\epsilon_{i\sigma}\}$ and $\mu$ are transformed according to
$\epsilon_{i\sigma} \rightarrow
\varepsilon_{i\sigma}\equiv\epsilon_{i\sigma}-\mu$ and 
$\mu \rightarrow \nu(\mu,\{\epsilon_{i\sigma}\})$.
Here,
$\nu_{\sigma}$ can be
any function such that
the Jacobian,
$
\left|
\frac{\partial (\nu,\{\varepsilon_{i\sigma}\})}
     {\partial (\mu,\{\epsilon_{i\sigma}\})}
\right|
$,
is non-zero. This is
because the relation
$\frac{\partial \dot{E}_{\rm KS}}{\partial \nu} = 0$ is identical to
equation \eqref{convcond}.
This means that the eigenvalues have meaning only in terms of their differences from
the reference energy $\mu$
 (or other eigenvalues,
$\epsilon_{i\sigma}-\epsilon_{j\sigma'}
=\epsilon_{i\sigma}-\mu + \mu-\epsilon_{j\sigma'}
$),
which is physically reasonable.

In some cases, eigenvalues appear in $\dot{E}_{\rm KS}$ in a symmetric
form
that is
invariant under constant shift, e.g.,
$\epsilon_{i\sigma}\rightarrow\epsilon_{i\sigma}+C_{\sigma}$,
for each $\sigma$.
This can be interpreted as the capability of having two independent
energy-scale references for $V_{{\rm eff},\uparrow}$ and
$V_{{\rm eff},\downarrow}$, separately.
In this case, $\mu_{\sigma}$ can be defined as a minimum real number 
that holds for
\begin{equation}
 N_{\sigma}
  =
  \sum_{i}
  \theta(\mu_{\sigma}-\epsilon_{i\sigma}),
\label{number-spin}
\end{equation}
and
$\mu$ in $\varepsilon_{i\sigma}$
should be replaced
with
$\varepsilon_{i\sigma} = \epsilon_{i\sigma} - \mu_\sigma$.
Then, $n_{\sigma}$ can be constructed from $V_{{\rm eff},\sigma}$ and 
$\mu_{\sigma}$ irrespective of the values for the opposite spin.
Let $\ddot{E}_{\rm KS}$ denote such a functional, with
\begin{equation}
 \ddot{E}_{\rm KS}
  [n_{\uparrow},n_{\downarrow}]
 =
 \ddot{E}_{\rm KS}
  [
    \{
       \{\psi^*_{i\sigma},\psi_{i\sigma}\}_i[n_{\sigma}],
       \{\varepsilon_{i\sigma}\}_i[n_{\sigma}]
    \}_{\sigma}
  ].
\end{equation}
A similar equation to equation \eqref{convcond}, i.e.,
\begin{equation}
  \frac
    {\partial \ddot{E}_{\rm KS}}
    {\partial \mu_{\sigma}}
 +
 \sum_{i\sigma}
 \frac
    {\partial \ddot{E}_{\rm KS}}
    {\partial \epsilon_{i\sigma}}
 =
 0,
\label{convcond2}
\end{equation}
always holds,
and the variation of $\ddot{E}_{\rm KS}$ with respect to $\mu_{\sigma}$
gives 
\begin{equation}
\sum_{i}
 \left[
  \delta (\mu_{\sigma}-\epsilon_{i\sigma})
   \int
   \left\{
  \xi+\mu_{\sigma}-V_{{\rm xc},\sigma}({\vec r}')
   \right\}
  \psi_{i\sigma}^*({\vec r}')
  \psi_{i\sigma}({\vec r}')
  d{\vec r}'
  \right] 
  +
    \frac
       {\partial \ddot{E}_{\rm xc}}
       {\partial \mu_{\sigma}}
  +
  \lambda
  \frac{\partial P}{\partial \mu_{\sigma}}
 =
 0.
\label{FAequation2}
\end{equation}
In this case, one can show the following equivalence:
$\eqref{ModSHTS}, \eqref{OnWayFA3}\Longleftrightarrow
\eqref{ModSHTS},\eqref{FAequation2}$.
Equation \eqref{FAequation2} is the same as our previously proposed
equation\cite{fukazawa2010new}, except in that case some terms were removed using
Lagrange multipliers.
Therefore, equation \eqref{OnWayFA3} can be regarded as an extended
version of the previous equation to a general $E_{\rm KS}$.

Let us return to the general discussion.
We obtained equations \eqref{NewMainEq} and \eqref{FAequation}
by minimizing $E_{\rm KS}$.
From the equivalences given in
\eqref{equiv0} and \eqref{equiv2}, the following relationship can be obtained
\begin{equation}
 \eqref{NewMainEq},  \eqref{FAequation}
 \Longleftrightarrow
 \eqref{ModSHTS}, \eqref{OnWayFA3}, \eqref{convcond}.
 \label{equiv1} 
\end{equation}
Therefore, we can expect that
\eqref{ModSHTS},
\eqref{OnWayFA3}, and \eqref{convcond}
on the right-hand side
contain full information on the extrema.
The first
equation, \eqref{ModSHTS}, is almost equivalent to the SHTS equation,
apart from the presence of the last term,
which comes from the optimization with respect to the eigenvalues.
In the case of the LDA or EXX, there is no contribution of eigenvalues
to $E_{\rm KS}$, apart from the $\delta$-functional level-crossing effect;
therefore, $R_{{\rm xc},i\sigma}=0$.
However, one can not always neglect the 
analytic eigenvalue-dependence of the functional
when it includes eigenvalues explicitly.
The Lagrange multiplier, $\lambda$, can be determined from equation
\eqref{Rsum}, which was obtained by integrating \eqref{ModSHTS}
with respect to ${\vec r}$. 
Recalling the existence of $\delta\mu$,
one can find
\begin{equation}
 \lambda \frac{\partial P}{\partial \epsilon_{i\sigma}}=
  \left\{
   \begin{array}{cc}
    -\frac{1}{{\mathcal N}_{\sigma}} \sum_j R_{{\rm xc},j\sigma}
     & (\mbox{if}\ \mu-\epsilon_{i\sigma}=0)\\
    0 &  (\mbox{otherwise})\\
   \end{array}
  \right.,
\end{equation}
where ${\mathcal N}_{\sigma}$ denotes the number of 
highest-occupied orbitals of spin state $\sigma$.
The exchange-correlation component of the effective potential,
$V_{{\rm xc},\sigma}$, is given by inverting 
$\mathbf{G}_{i\sigma}({\vec r}', {\vec r})\psi_{i\sigma}({\vec r})
+{\rm c.c.}$, however, the unphysical invariance under
\eqref{unPhysFreedom} of $V_{{\rm xc},\sigma}$ 
due to the irregularity of
$\mathbf{G}_{i\sigma}$
emerges here. 

As mentioned above,
the second equation, \eqref{OnWayFA3}, is a generalized version of 
the equation used
in our previous
paper\cite{fukazawa2010new}
to correct for the unphysical degree of freedom.
It can also be related to
a prescription given by KLI\cite{KLI:PhysRevA.45.101}
for setting the constants,
which
becomes identical to our method
for some classes of approximated Kohn-Sham
functionals
when there is no degeneracy in the system\cite{fukazawa2010new}.
Whereas those authors were required to consider a region in which the highest occupied
orbitals were
dominant, which restricted the validity of their
discussion
to isolated systems, it is not necessary to assume the existence of such
a region to
derive equation \eqref{OnWayFA3}.
Nevertheless, this equation manifests itself as regards the influence of the effective potential
on the highest occupied orbitals of the system, and reduces the
degree of freedom allowed in \eqref{ModSHTS} from two to one.

The last equation, \eqref{convcond}, is a requirement for $E_{\rm KS}$ to 
behave as a density functional around the extrema, which is also
mentioned above. We emphasize again that this equation always holds
for density functionals that do not depend on the choice of 
energy-scale reference. In this case, the degree of freedom,
which
remains even after consideration of equation \eqref{OnWayFA3}, is not
required to be fixed, because the addition of a constant to both
$V_{{\rm eff},\uparrow}$ and $V_{{\rm eff},\downarrow}$ must have 
no effect on the physical properties.
If $E_{\rm KS}$ depends on the reference energy, this constant has an
effect and is set by equation \eqref{convcond}.

\subsection{Static RPA correlation}
We discuss an application of the reformulated method
to a calculation with static RPA functionals.
Let us first provide some notation and definitions used in the below discussion,
for convenience.
First, we define some operations for functions of two sets of time-space
variables.
The product, $f\cdot g$, of functions $f(x_1,x_2)$ and $g(x_1,x_2)$
is defined as
\begin{equation}
 f\cdot g (x_1,x_2)
  =
  \int dx'
  f(x_1,x')
  \,
  g(x',x_2),
\end{equation}
where the $x$'s
denote sets of time-space variables such that $x_1=({\vec x}_1, t_1)$.
The identity, $1(x,x')$, must be the delta function, $\delta(x-x')$.
If a function, $f$, has an inverse, we
denote it by $f^{-1}$, with
$f\cdot f^{-1}=f^{-1}\cdot f=1$.
Some functions have time dependence but depend only 
on the difference in the two time-variables.
In this case, 
we use the expression,
$f(x,x')=f({\vec r},{\vec r}', t-t')$.
It is also convenient to define an operator, $\Tr^{0}$,
which is the trace
of the space-variables, such that
\begin{equation}
 \Tr^{0}[f]
  =
  \int d{\vec r} \,
  f({\vec r},{\vec r},0).
\end{equation}
This type of trace of a product of two or more
functions is invariant under cyclic permutation of the order of the 
product,
because
$
\Tr^0[f\cdot g]
=
\int d{\vec r} d{\vec r}' d\tau
f({\vec r}, {\vec r}',t-\tau)
g({\vec r}',{\vec r} ,\tau-t)
=
\int d{\vec r} d{\vec r}' d\tau'
g({\vec r}',{\vec r} ,t-\tau')
f({\vec r}, {\vec r}',\tau'-t)
=
\Tr^0[g\cdot f],
$
with
$\tau'=-\tau+2t$.
Second, 
we let tilde denote the Fourier transform of a function with respect to
time.
In this case, the product becomes
$
\tilde{f}\cdot\tilde{g}({\vec r}_1,{\vec r}_2,\omega)
=
\int d{\vec r}'
\tilde{f}({\vec r}_1,{\vec r}',\omega)
\,
\tilde{g}({\vec r}',{\vec r}_2,\omega),
$
and the Fourier transform of the trace is
$
 \tilde{\Tr}^0[\tilde{f}]
  =
  \frac{1}{2\pi}
  \int d\omega
  \int d{\vec r}
  \tilde{f}({\vec r},{\vec r},\omega).
$

Finally, we introduce some notation.
The Coulomb interaction is denoted by $v$,
which can be written as
$
 v(x, x')
  = 2\,
    {\delta(t-t')}
    /
    \,
    {\left|{\vec r}-{\vec r}'\right|}
$
in atomic Rydberg units,
and transformed to
$
 \tilde{v}({\vec r}, {\vec r}', \omega)
  =  \tilde{v}({\vec r}, {\vec r}')
  = 2
    /
    {\left|{\vec r}-{\vec r}'\right|}.
$
We also use the density matrix,
$
 \tilde{\rho}_{\sigma}({\vec r},{\vec r}', \omega)
  =
  \sum_{i}
  \delta(\omega-\epsilon_{i\sigma})
  \psi_{i\sigma}({\vec r})
  \psi^*_{i\sigma}({\vec r}').
$
The causal Green's function of the auxiliary system is denoted by
$G_{\sigma}(x,x')$, and the retarded Green's function
by $G^{\rm R}_{\sigma}(x,x')$.
The causal Green's function is important as a building block in
the formulation, while the retarded Green's function is convenient for
actual implementation to numerical procedures.

The application to the EXX functional,
\begin{equation}
 {E}^{\rm EXX}
  =
  \int^{\mu_\sigma}_{-\infty} dE
  \int^{\mu_\sigma}_{-\infty} dE'
  \int d{\vec r}
  \int d{\vec r}'
  \tilde{\rho}({\vec r}, {\vec r}',E )
  \,
  \tilde{v}({\vec r}, {\vec r}')
  \,
  \tilde{\rho}({\vec r}',{\vec r} ,E'),
\end{equation}
has already been discussed in our previous paper\cite{fukazawa2010new},
based on the
the SHTS equation and
equation \eqref{FAequation2}.
This treatment happens to be in accordance with the present
formalism,
because 
$E^{\rm EXX}$ has the symmetry required in $\ddot{E}_{\rm KS}$
and ${R}_{{\rm xc},i\sigma}$
of EXX is zero.
However, this is not true in the case of the 
correlation-functional with the random-phase-approximation,
${E}_{\rm c}^{\rm RPA}$,
which is defined as
\begin{equation}
 {E}_{\rm c}^{\rm RPA}
  =
  -
  \Tr^{0}[\log(1-v\cdot \Pi)+v\cdot \Pi].
\end{equation}
Here, the logarithm is defined by its series expansion and $\Pi$ is 
the ring polarization insertion, where
\begin{equation}
 \Pi(x,x')
  =
  \sum_{\sigma}
  \Pi_{\sigma}(x,x')
  =
  -i
  \sum_{\sigma}
  G_{\sigma}(x,x')
  \,
  G_{\sigma}(x',x).
\end{equation}
A convenient feature of ${E}^{\rm RPA}_{\rm c}$ is that 
its variation can be attributed to that of
$\Pi$, such that
\begin{equation}
 \delta {E}_{\rm c}^{\rm RPA}
  =
  \frac{i}{2}
  \Tr^{0}[W \cdot \delta \Pi],
  \label{dERPA}
\end{equation}
where $W$ is defined as
\begin{equation}
 W=(1-v\cdot \Pi)^{-1}\cdot v\cdot \Pi \cdot v.
\end{equation}
Therefore, 
${R}_{{\rm xc},j\sigma}$ for the RPA correlation functional
can be written as
\begin{equation}
 {R}_{{\rm c},j\sigma}^{\rm RPA}
  =
  \frac{i}{2}
  \tilde{\Tr}^0
  \left[
   \tilde{W}
   \cdot
   \left(
   \frac
     {\partial \tilde{\Pi}}
     {\partial \epsilon_{j\sigma}}
   \right)_{\rm Rem.}
   \right],
\end{equation}
where the subscript ``Rem.'' denotes the non-delta functional component
forming $R_{{\rm xc},i\sigma}$ in equation \eqref{DecomEe}.
Note that ${R}_{{\rm c},j\sigma}^{\rm RPA}$ cannot be ignored
in this case, because 
\begin{eqnarray}
 \left(
  \frac
    {\partial \tilde{\Pi}_{\sigma}}
    {\partial \epsilon_{j\sigma}}
 \right)_{\rm Rem.}
\nonumber \\
 =
 \theta(\epsilon_{j\sigma}-\mu_{\sigma})
 \,
 \psi  _{j\sigma}({\vec r})
 \psi^*_{j\sigma}({\vec r}')
 \nonumber \\
  \qquad
 \times
 \int^{\mu_{\sigma}}_{-\infty} dE\,
 \tilde{\rho_{\sigma}}({\vec r}',{\vec r},E)
 \left\{
   \frac
     {1}
     {(\omega-(\epsilon_{j\sigma}-E)+i\eta)^2}
   +
   \frac
     {1}
     {(\omega-(E-\epsilon_{j\sigma})-i\eta)^2}
 \right\}
 \nonumber \\
 -
 \theta(\mu_{\sigma}-\epsilon_{j\sigma})
 \,
 \psi  _{j\sigma}({\vec r}')
 \psi^*_{j\sigma}({\vec r})
 \nonumber \\
  \qquad
 \times
 \int^{\infty}_{\mu_{\sigma}} dE\,
 \tilde{\rho_{\sigma}}({\vec r},{\vec r}',E)
 \left\{
   \frac
     {1}
     {(\omega-(E-\epsilon_{j\sigma})+i\eta)^2}
   +
   \frac
     {1}
     {(\omega-(\epsilon_{j\sigma}-E)-i\eta)^2}
 \right\},
 \nonumber \\
\end{eqnarray}
is not zero in general. This is a natural consequence from the explicit 
dependence of ${E}^{\rm RPA}_{\rm c}$ on eigenvalues.
Instead of applying the RPA correlation functional,
the static approximation is utilized in the following calculations
in order to reduce the computational cost,
following Kotani\cite{kotani98}.
This is justified when $\tilde{W}$ is slowly-varying with respect to
the frequency, $\omega$, such that $W$ can be efficiently approximated by
\begin{equation}
 W(x,x')
  =
  \frac{1}{2\pi}
  \int d\omega\,
  \tilde{W}({\vec r},{\vec r}',\omega)
  e^{-i\omega(t-t')}
  \sim
  \tilde{W}({\vec r},{\vec r}',0)\,
  \delta(t-t').
\end{equation}
Then, equation \eqref{dERPA} becomes
\begin{equation}
 \delta {E}_{\rm c}^{\rm sRPA}
  =
  \frac{i}{2}
  \Tr^{0}
  \left[
   \int d{\vec r}'\,\tilde{W}({\vec r},{\vec r}',0)
   \,
   \delta \Pi({\vec r}',{\vec r},0)
  \right].
  \label{deltaEsRPA}
\end{equation}
With 
this approximation, ${R}_{{\rm c},i\sigma}$ vanishes, because 
$\Pi({\vec r}',{\vec r},0)$ does not explicitly depend on 
eigenvalues, apart from the stepwise dependency of 
$\theta(\mu-\epsilon_{i\sigma})$.
This feature of the static approximation supports the validity of
Kotani's
calculation, even though the existence of the correction term has not been 
recognized.

The reduction of the computational cost is achieved 
in the static approximation by avoiding
integrations with respect to frequency.
However, when obtaining
$\tilde{W}({\vec r}, {\vec r}',0)$, the calculation of
\begin{equation}
 \tilde{\Pi}_{\sigma}({\vec r},{\vec r}',0)
  =
  2
  \int_{-\infty}^{\mu_\sigma} dE'
    \,
    \tilde{\rho}_{\sigma}({\vec r}',{\vec r} ,E')
  \int^{\infty} _{\mu_\sigma} dE
    \,
    \frac
    {
      \tilde{\rho}_{\sigma}({\vec r} ,{\vec r}',E )
    }
    {
      E'-E+i\eta
    },
    \label{directPi}
\end{equation}
is still time consuming.
Here, $\eta$ is a real positive infinitesimal.
A considerable obstacle to a fast and precise calculation
is that $\tilde{\Pi}_{\sigma}({\vec r},{\vec r}',0)$
involves summation over all the unoccupied orbitals.
In the previous calculations, an energy cut-off was introduced in the 
integration with respect to $E$ in equation \eqref{directPi}.
This may create a question regarding precision, or require 
additional computational time even when the precision can be
assured using numerical procedures.
A significantly improved scheme is obtained by rewriting the integral as
\begin{eqnarray}
 \int^{\infty} _{\mu_\sigma} dE
  \,
  \frac
  {
    \tilde{\rho}_{\sigma}({\vec r} ,{\vec r}',E )
  }
  {
    E'-E+i\eta
  }
  =
  \tilde{G}^{\rm R}_{\sigma}({\vec r},{\vec r}',E')
  -
 \int_{-\infty}^{\mu_\sigma} dE
  \,
  \frac
  {
    \tilde{\rho}_{\sigma}({\vec r} ,{\vec r}',E )
  }
  {
    E'-E+i\eta
  },
\\
 =
  \theta(E'-\mu_\sigma)
  \tilde{G}^{\rm R}_{\sigma}({\vec r},{\vec r}',E')
  -
  \frac{1}{\pi}
  \Im
 \int_{-\infty}^{\mu_\sigma} dE
  \,
  \frac
  {
    \tilde{G}^{\rm R}_{\sigma}({\vec r},{\vec r}',E)
  }
  {
    E'-E+i\eta
  },
\end{eqnarray}
which follows from
$
\Im\left[z/(\alpha+i\beta)\right]
=
(\Im z)/(\alpha-i\beta)
-
z\beta/(\alpha^2+\beta^2)
$,
where $\alpha$ and $\beta$ are real numbers.
The resultant expression of
$\tilde{\Pi}_{\sigma}({\vec r},{\vec r}',0)$ is 
\begin{equation}
 \tilde{\Pi}_{\sigma}({\vec r},{\vec r}',0)
  =
  -
  \frac{2}{\pi}
  \int_{-\infty}^{\mu_\sigma} dE'
    \,
    \tilde{\rho}_{\sigma}({\vec r}',{\vec r} ,E')
  \,
  \left[
  \Im
  \int_{-\infty} ^{\mu_\sigma} dE
    \,
    \frac
    {
      \tilde{G}^{\rm R}_{\sigma}({\vec r} ,{\vec r}',E )
    }
    {
      E'-E+i\eta
    }
  \right].
  \label{tildeD2}
\end{equation}
However, 
this integral is only justifiable
by 
infinitesimally shifting
the path of the integral with respect to $E$ upward
in the complex plane,
because, otherwise,
the integral 
diverges due to the poles of second order
in 
$
    \frac
    {
      \tilde{G}^{\rm R}_{\sigma}({\vec r} ,{\vec r}',E )
    }
    {
      E'-E+i\eta
    }.
$
Even with such a manipulation, 
the existence of the poles, which are of nearly second order,
prevents numerical integration of \eqref{tildeD2}.

However, we found that this concept can be applied in practical calculations
by performing an analytical continuation on the integrand of \eqref{tildeD2} and 
separating the paths of both integrals from each other, 
because the second order nature of the poles comes from the closeness
between these two paths.
Such an analytic continuation is obtained
from
an analytic continuation of the square bracket given by
\begin{equation}
 \frac{1}{2i}
  \left[
   \int_{C_1}
     dE
     \,
     \frac
     {
       \tilde{G}^{\rm R}_{\sigma}({\vec r},{\vec r}',E)
     }
     {
       E'-E+i\eta
     }
   -
   \int_{C_2}
     dE
     \,
     \frac
     {
       (\tilde{G}^{\rm R}_{\sigma}({\vec r},{\vec r}',E^*))^*
     }
     {
       E'-E+i\eta
     }
  \right],
\end{equation}
where $C_1$ is a path in the upper-half plane that begins at $-\infty$
and ends at $\mu_\sigma$, while $C_2$ shares the same ends as $C_1$ but
lies in the lower-half plane. Since
this expression is also analytic with respect to $E'$,
the path for the integral with respect to $E'$ can also be deformed.
Then, a preferable form of 
$\tilde{\Pi}_{\sigma}({\vec r},{\vec r}',0)$ is obtained 
as
\begin{eqnarray}
 \tilde{\Pi}_{\sigma}({\vec r},{\vec r}',0) 
 \nonumber \\
  =
  \frac{1}{\pi}
  \Re
  \left[
   \int_{C_3}
     dE'
     \tilde{G}^{\rm R}_{\sigma}({\vec r}',{\vec r},E')
    \left\{
     \int_{C_1}
       dE
       \,
       \frac
       {
         \tilde{G}^{\rm R}_{\sigma}({\vec r},{\vec r}',E)
       }
       {
         E'-E+i\eta
       }
     -
     \int_{C_2}
       dE
       \,
       \frac
       {
         (\tilde{G}^{\rm R}_{\sigma}({\vec r},{\vec r}',E^*))^*
       }
       {
         E'-E+i\eta
       }
  \right\}
  \right],
  \nonumber \\
\label{finalPi}
\end{eqnarray}
where $C_3$ is a path that begins at $-\infty$, ends at
$\mu_\sigma$, and lies in the region enclosed by $C_1$ and the real
axis.
In the following calculations, we used this expression, which is 
free from the errors originating from the energy cut-off
in equation \eqref{directPi}.

\section{Numerical examples}
\label{SS.Results}
In this section, we show results obtained
from our implementation of the EXX+sRPA method 
described in the previous section.
In order to achieve high calculational speed, we have adopted
the modified version\cite{fukazawa2010new} of
the Krieger-Li-Iafrate (KLI)
approximation\cite{Krieger1990256,KLI:PhysRevA.45.101}
to solve equations \eqref{FAequation} and \eqref{ModSHTS}.
Although we recognize that a significant number of studies have been conducted on the development
of approximations\cite{staroverov2006optimized,staroverov2006effective,
izmaylov2007effective,izmaylov2007self,PhysRevA.64.042506,
della2001efficient}
beyond the KLI, or on acceleration schemes\cite{PhysRevLett.90.043004,
PhysRevB.68.035103,
kosaka2006simple,PhysRevB.85.235126}
to 
solve the SHTS equation directly, 
we chose the KLI approximation as a first trial step toward a 
practical and accurate method.
In order to solve the single-particle problem,
we used 
the Korringa-Kohn-Rostoker (KKR) Green's function method,
because
it can calculate the
retarded Green's function of the auxiliary system
appearing in \eqref{finalPi} directly. 
The path, $C_2$, in \eqref{finalPi} can be 
chosen so that $C^*_2$ becomes identical to $C_1$ or $C_3$.
The retarded Green's functions
on $C_1$ and $C_3$
were constructed in our calculations
with $C^*_2=C_1$.
We included all the contributions of core orbitals,
and all the core orbitals are self-consistently determined.

\begin{figure}[hbtp]
 \begin{center}
  \includegraphics[width=50mm,angle=-90]{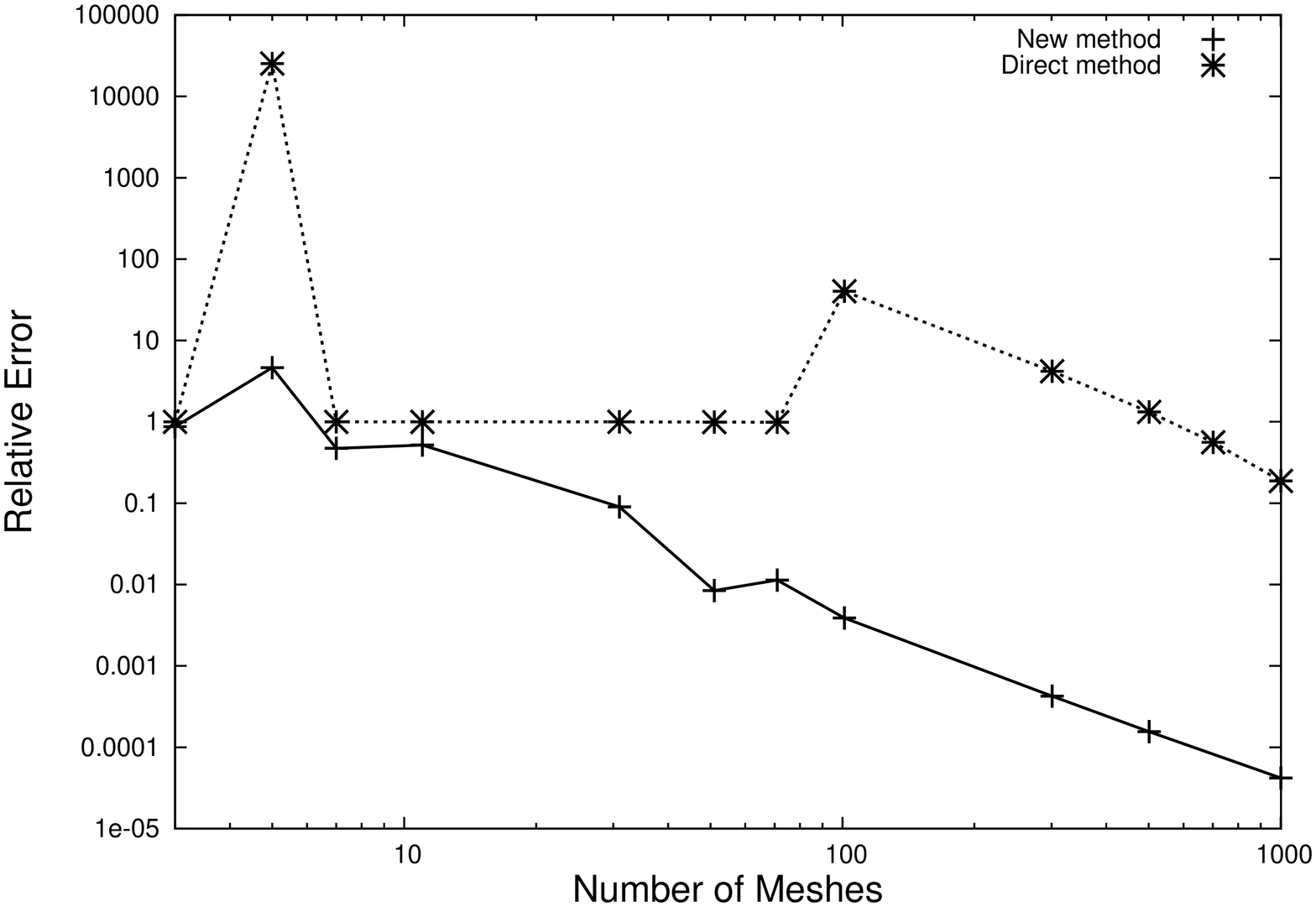}
  \includegraphics[width=50mm,angle=-90]{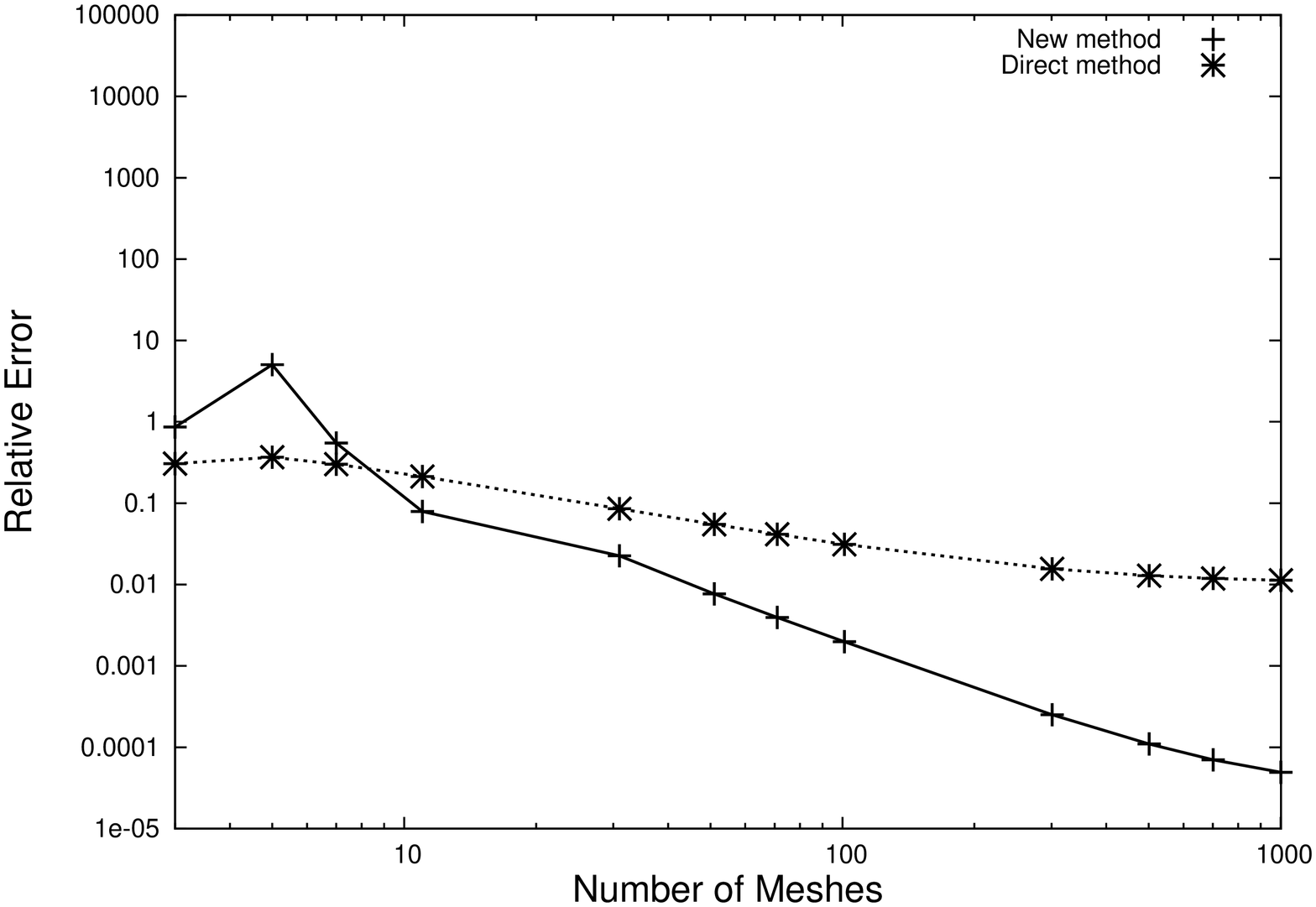}
  \caption{    
    Relative errors   of calculated
    polarization insertion 
    versus number of meshes for
    numerical integration (log-log plots).
    In the calculation, 
    the model Green's
    functions, $G^{\rm I}$ (left) and  $G^{\rm II}$ (right), were used.
  \label{RPAtestcomp}}
 \end{center}
\end{figure}

Before proceeding to realistic systems, we performed test calculations
using \eqref{finalPi}
and compared the results to those from the direct calculation given by
\eqref{directPi},
with
$
\tilde{\rho}({\vec r},{\vec r}',E)
=
-\frac{1}{\pi}
\Im \tilde{G}^{\rm R}({\vec r},{\vec r}',E)
$.
Since only the energy dependence of the retarded Green's
function is important for our purposes,
we used the following two-model Green's functions,
which depend on the energy only:
\begin{equation}
 G^{\rm I}(E)
  =
  \frac{1}{E+A+i\eta}
  +
  \frac{1}{E-A+i\eta},
\end{equation}
and 
\begin{equation}
 G^{\rm II}(E)
  =
  \log\left(\frac{E+B+C}{E-B+C}\right)
  +
  \log\left(\frac{E-B+C}{E-B-C}\right),
\end{equation}
with $A=0.5$, $B=0.5$, and $C=0.55$,
and $\mu$ set to zero.
$G^{\rm I}$ has two $\delta$-functions in its
imaginary component, representing a rapidly-varying function,
while $G^{\rm II}$ has two
rectangles in its imaginary component, representing a slowly-varying
function.
The direct method is applied with the paths of the integral shifted
upward by $10^{-3}$ in the calculation, in order to obtain converging results with
a finite number of meshes.
As shown
in Figure \ref{RPAtestcomp}, in the direct method, 
the error of the integrals decreased slowly as number
of meshes increased.
On the other hand, the convergence
was much improved in our new method
for both the model Green's functions.

Next, we apply these techniques to realistic calculations involving bcc-Fe,
fcc-Co, and fcc-Ni.
These magnetic metals are expected to constitute stringent test cases because
they have been proven to be
sensitive to the treatment of the unphysical degree of freedom in 
\eqref{ModSHTS}, and also to correlation effects included in the
calculation.
Table \ref{momenttable} shows the comparison of
the predicted magnetic moments obtained by various methods.
It can be seen from this table that the magnetic moments obtained by the EXX
calculations, which were performed without any correlation functional,
are reduced significantly by taking 
the correlation with the static RPA into account.
The deviation of the results of our EXX+sRPA method from the experimental values
are comparable to those of the LDA calculations.
In this sense, use of the modified KLI method does not significantly
reduce the accuracy of description of the system.
On the other hand, 
the values reported by Kotani\cite{kotani98} exhibited closer agreement with the
experimental values than our results.
The systematic underestimation of the magnetic moments in our results
could be explained as being due to the effects of unoccupied orbitals, which were 
previously neglected in the calculation of equation \eqref{directPi}.

\begin{table}[htb]
 \begin{center}
  \begin{tabular}{cccccc}
   \hline
   ($\mu_{\rm B}$)
   &EXX+sRPA& EXX &  LDA
   & Exp.
   \cite{danan1968new,besnus1970magnetic,reck1969orbital}\\
   \hline
   Fe & 1.93 & 3.38 & 2.28 & 2.12 \\
   Co & 1.47 & 2.26 & 1.60 & 1.59 \\
   Ni & 0.54 & 0.82 & 0.59 & 0.56 \\
   \hline
  \end{tabular}
 \end{center}
 \caption{Predicted spin moments of the ferromagnetic Fe, Co, and Ni
 obtained using the KKR-OEP-KLI technique,
 and those from experiments.
 The experimental values are deduced from the total magnetic moments
 from Ref. \cite{danan1968new} and \cite{besnus1970magnetic}
 and the ratios of the spin contribution from
 Ref. \cite{reck1969orbital}.
 \label{momenttable}
 } 
\end{table}

\begin{figure}[htbp]
 \begin{center}
  \includegraphics[width=7cm]{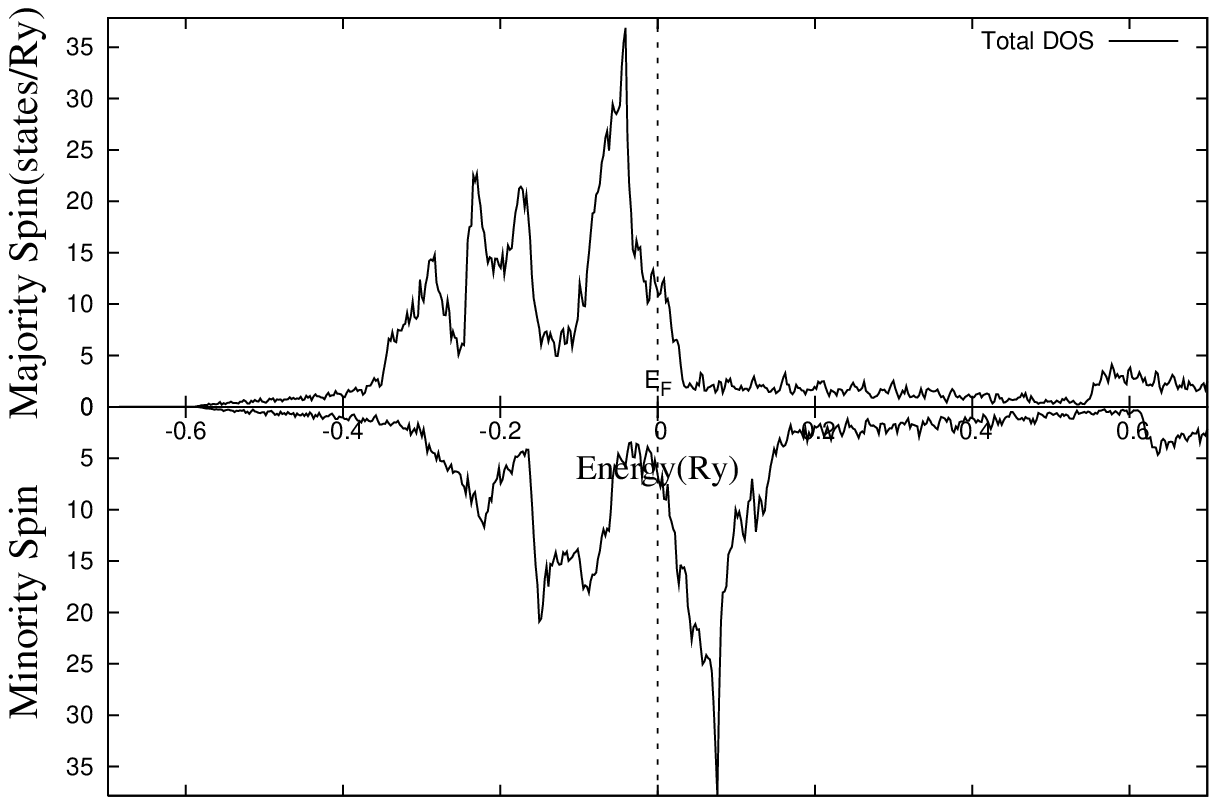}
  \includegraphics[width=7cm]{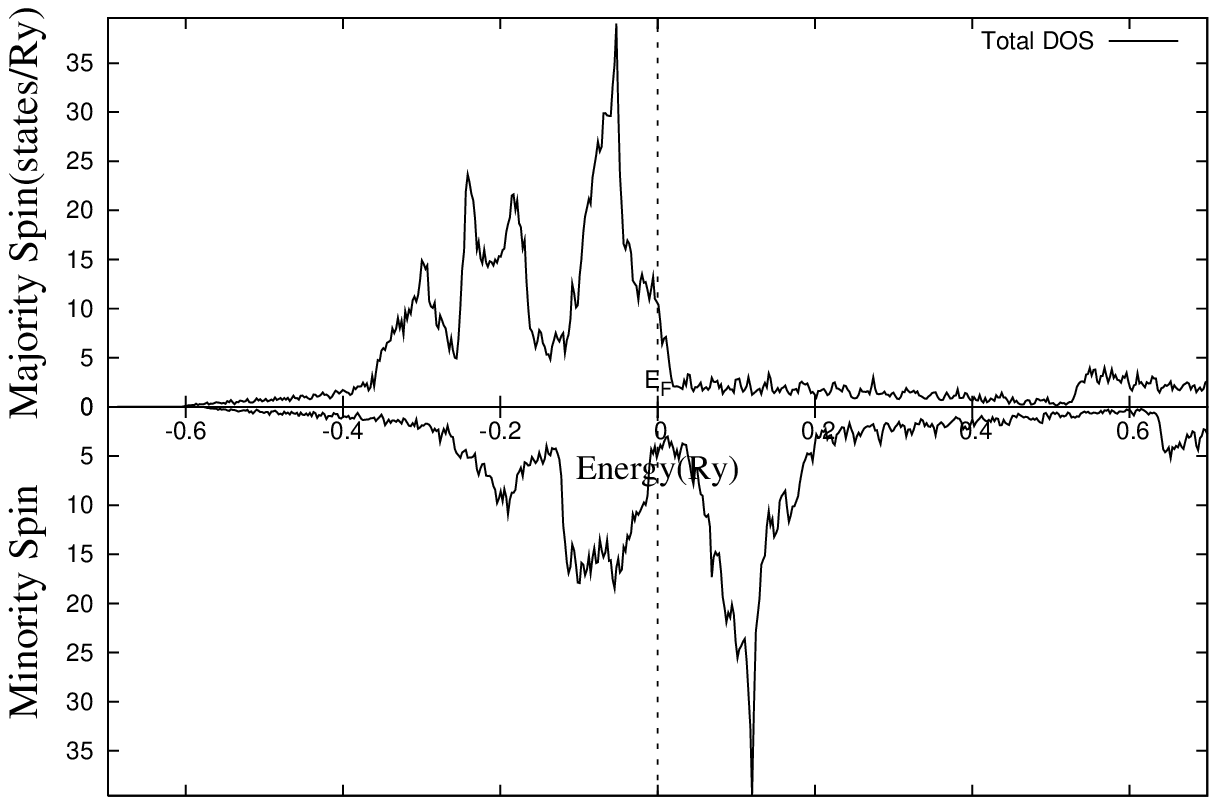}
 \end{center}
 \caption{\label{DOS_Fe} Total DOS of Fe
 calculated
 with (left) EXX+sRPA
 and (right) LDA.
 }
\end{figure}

\begin{figure}[htbp]
 \begin{center}
  \includegraphics[width=10cm,angle=0]{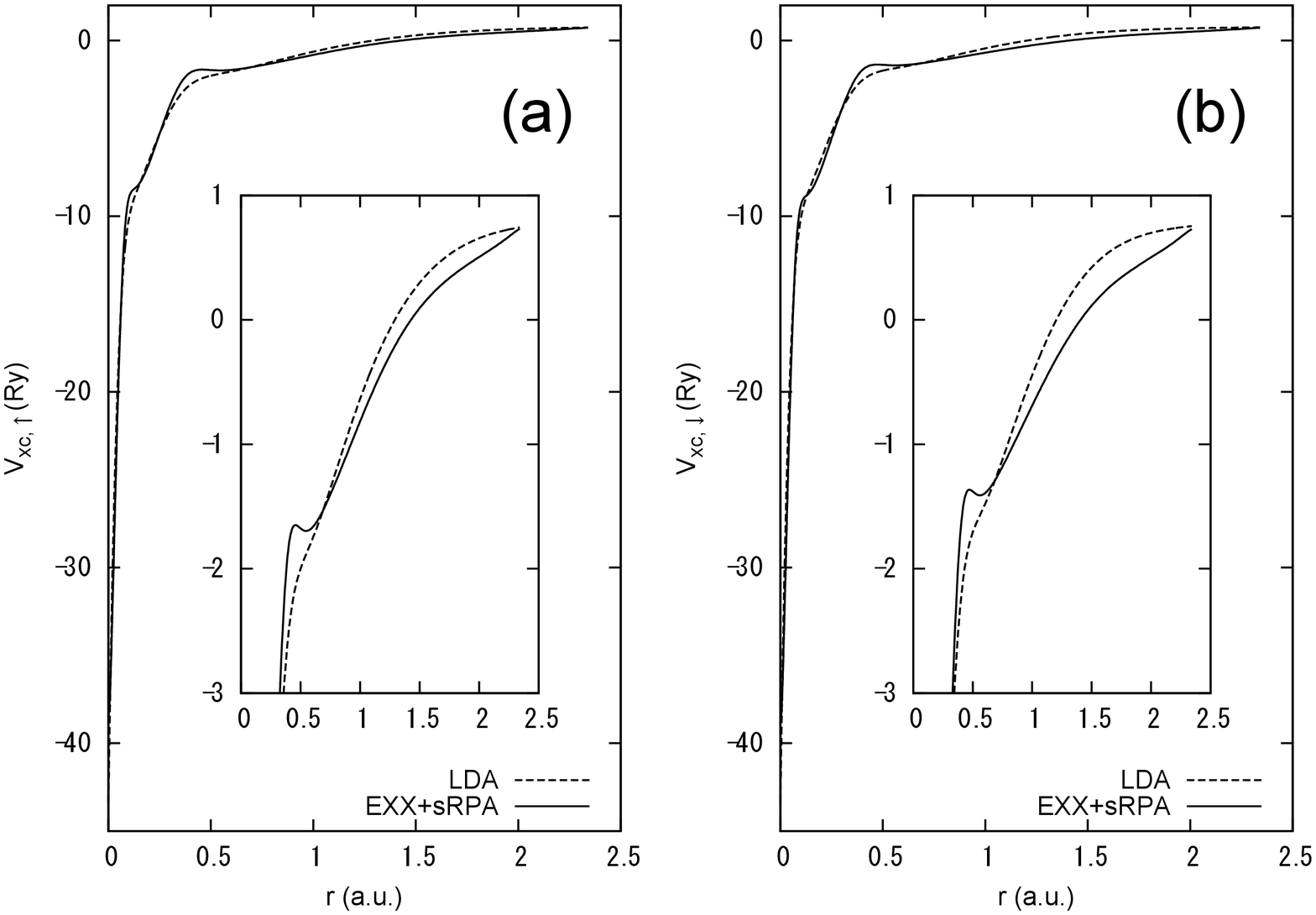}
  \includegraphics[width=10cm,angle=0]{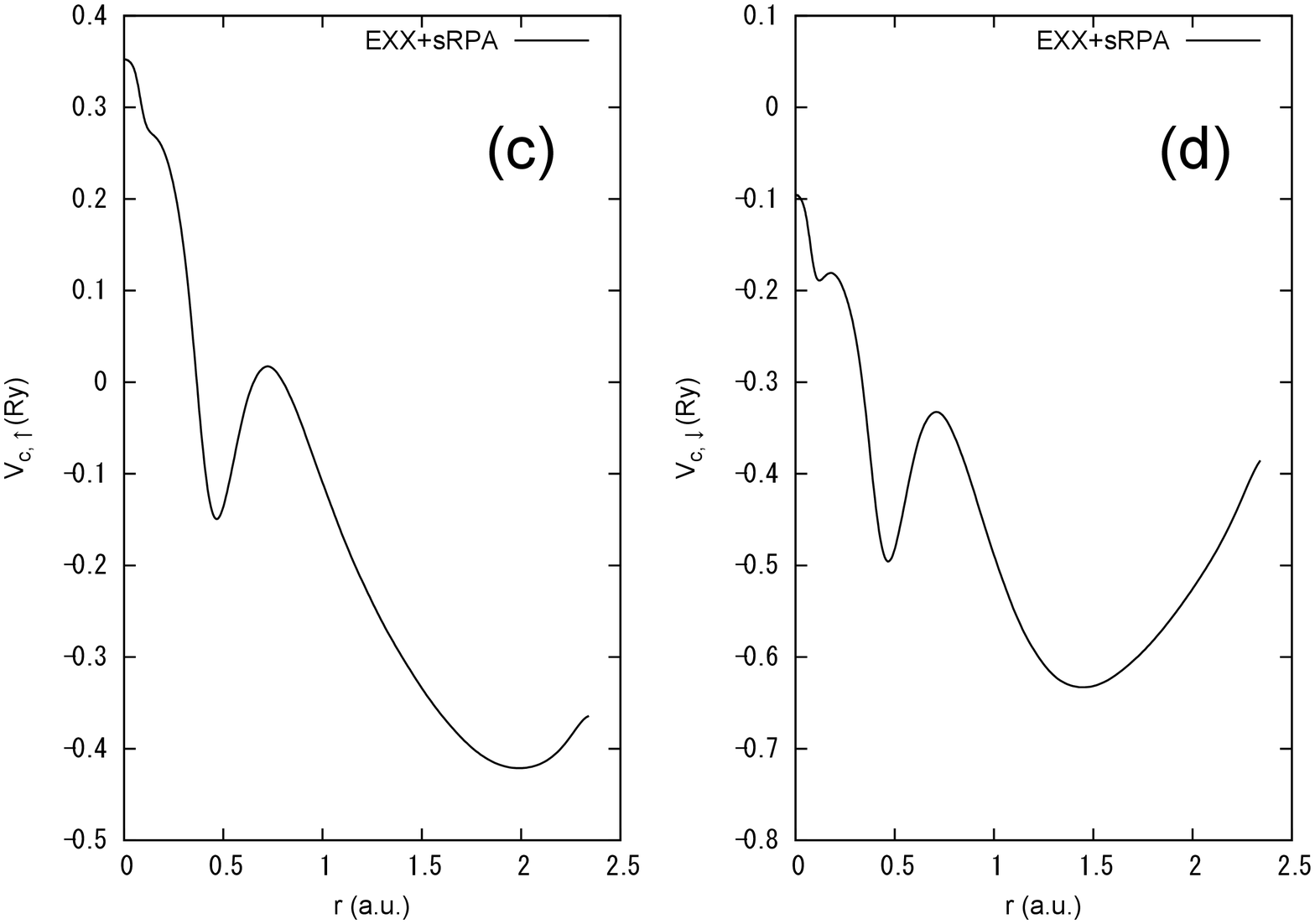}
 \end{center}
 \caption{\label{pot_Fe}
 Exchange-correlation potentials of bcc-Fe for the EXX+sRPA
 (solid line)
 and the LDA (broken line) techniques with (a) majority spin and (b) minority spin.
 The insets are magnifications of the  external figures.
(c) and (d) are related to the correlation component of the potentials
 calculated for the EXX+sRPA method with (c)  majority spin and (d) minority spin. }
\end{figure}

As previously reported for an EXX case\cite{fukazawa2010new},
the DOS calculated using the EXX approach differs significantly from that of
the LDA due to the overestimated exchange splitting.
This behavior is significantly improved by taking the screening
effect within the static RPA correlation into account, as shown in
in Figure \ref{DOS_Fe}.
Despite the similarity in the DOS,
the resultant effective potentials seem rather different from 
each other. Figure \ref{pot_Fe} (a) and (b) compare the
EXX+sRPA exchange-correlation potentials for bcc-Fe with those of the
LDA.
While the LDA potentials have smooth shapes, 
those of the EXX+sRPA show typical dips, which correspond
to the exchange hole. These dips are compensated for to some extent by
the correlation components of the $V_{{\rm c},\sigma}$ potentials,
as shown in Figure \ref{pot_Fe} (c) and (d).

The time consumption of the calculations was approximately 10 s
per iteration with one CPU core, for which 14 GFLOPS
is expected at peak performance.
The total calculation time, beginning with the
effective atomic potentials calculated with the LDA,
was typically 6 h.
The values of the Green's function were generated at 598
points in the energy plane, and 145 points in the k-space
were used.

\section{Conclusion}
\label{SS.Conclusion}

We have presented a reformulation of the OEP method. Some theoretical
limitations existing 
in conventional derivations have been overcome by carefully considering the dependency
of the energy functional on
eigenvalues.
Two equations, \eqref{ModSHTS} and \eqref{OnWayFA3}, are particularly important.
Equation \eqref{ModSHTS} suggests the existence of a correction term to
the SHTS equation
arising from a dependency 
of the energy functional on eigenvalues, 
which was neglected previously.
Although the use of conventional equations could be justified
for some exchange-correlation functionals,
as we demonstrated for the
LDA, EXX, and EXX+sRPA techniques in the present paper, 
we also showed that the RPA correlation functional offers a
counterexample.
Equation \eqref{OnWayFA3} provides a method of 
eliminating an unphysical degree of freedom that could be permitted in equation
\eqref{ModSHTS}, which can be regarded as a generalized form of 
our previous equation\cite{fukazawa2010new}, or the prescription proposed
by KLI\cite{KLI:PhysRevA.45.101}.

An efficient method of constructing a correlation potential within 
the EXX+sRPA approach has also been proposed. 
The advantage of this technique was confirmed by calculations
using the model Green's functions.
We also performed realistic calculations
of some transition metals, which are considered to constitute rigid test cases.
In order to accelerate the calculations,
we exploited
the modified\cite{fukazawa2010new} KLI
approximation\cite{Krieger1990256,KLI:PhysRevA.45.101}.
Although application of the modified KLI approximation would
introduce some errors to the calculation, 
the deviations of the results from the
experimental data were acceptable.
Therefore, the modified KLI approximation
combined with EXX+sRPA functionals
is regarded as an
adequate approximation for these systems at least,
provided a precise technique to manage the sRPA functional 
is exploited.

\section*{Acknowledgments}
The present study was partly supported by the Next Generation Super
Computing Project, MEXT, the Advanced Low Carbon Technology Research and
Development Program, JST, the Strategic Programs for Innovative Research
(SPIRE), MEXT, the Computational Materials Science Initiative (CMSI),
Japan, and by the outsourcing project of MEXT, the Elements Strategy
Initiative Center for Magnetic Materials (ESICMM), Japan. One of the
authors (T.F.) is grateful for the support of the Global COE program,
``Core Research and Engineering of Advanced Materials --
Interdisciplinary Education Center for Materials Science'', MEXT.

\bibliography{oep,kkr,dft,misc}

\end{document}